\documentclass[twocolumn,showpacs,preprintnumbers,amsmath,amssymb]{revtex4}

\usepackage{graphicx}
\usepackage{dcolumn}
\usepackage{bm}

\newcommand\cZ{{\mathcal Z}}

\begin{document}

\title{Queueing process with excluded-volume effect}
\author{Chikashi Arita}
\email{airta@math.kyushu-u.ac.jp}
\affiliation{Faculty of Mathematics, Kyushu University, Fukuoka 819-0395 Japan}

\begin{abstract}
We introduce an extension of the M/M/1 queueing process
 with a spatial structure and excluded-volume effect.
The rule of particle hopping is the same as for the 
 totally asymmetric simple exclusion process (TASEP).
A stationary-state solution is constructed
 in a slightly arranged matrix product form of the open TASEP.
We obtain the critical line that separates
 the parameter space depending on whether the model has the stationary state.
We calculate the average length of the model and the number of particles
 and show the monotonicity of the probability of the length in the stationary state.
We also consider a generalization of the model with backward hopping of particles allowed
 and an alternate joined system of the M/M/1 queueing process and the open TASEP.
\end{abstract}



\maketitle

\section{introduction}

The queueing process is a typical example of a Markov process \cite{R}.
One of the simplest queueing processes is of the so-called M/M/1 type,
 where the arrival of customers and their service obey the Poisson point process.
This model's stationary state is the geometric distribution
 that varies with the ratio of the arrival rate to the service rate.
The M/M/1 queueing process has no spatial structure
 and particles do not interact with each other.

On the other hand, the asymmetric simple exclusion process (ASEP)
 on a one dimensional lattice is one of the simplest Markov processes with interacting particles \cite{L}.
In the ASEP, each site can be occupied by at most one particle
 and each particle can hop to a nearest-neighbor site if it is empty.
The ASEP admits exact analyses of non-equilibrium properties by the matrix product ansatz 
 and the Bethe ansatz \cite{Sc}.
The matrix product form of the stationary state
 was firstly found in the totally ASEP
 with open boundaries (open TASEP), where each particle enters at the left end,
 hops forward (rightward) in the bulk and exits at the right end \cite{DEHP}.
Similar results have been obtained in
 various generalized ASEPs and similar models in one dimension
 with both open and periodic boundary conditions \cite{BE}. 

In this paper,
 we introduce an extension of the M/M/1 queueing process
 on a semi-infinite chain
 with the excluded-volume effect (hard-core repulsion)
 as in the open TASEP.
Each particle enters the chain
 at the left site next to the leftmost occupied site,
 hops and exits following the same rule as for the open TASEP.
A stationary-state solution is given by a slightly arranged matrix product form of the open TASEP.
The normalization constant is given by the generating function of that of the open TASEP.

This paper is organized as follows.
In Section 2, we briefly review
 the M/M/1 queueing process and the open TASEP.
In Section 3, we define the model.
In Section 4, we find a stationary state of the model.
This will be constructed
 in a slightly modified matrix product form
 of the open TASEP.
We obtain the critical line
 which separates the parameter space into the regions with and without the stationary state.
The critical line will be written
 in terms of the stationary current of the open TASEP.
We also calculate the average length of the system and the average number of particles
 on the assumption of the uniqueness of the stationary state.
We also show the monotonicity
 of the probability of the length (i.e. the position of the leftmost particle).
In Section 5, we generalize the model by allowing particles to hop backward.
It is fair to say that almost every calculation in
 Section 4 and 5 will be performed by using known formulae
 in studies of the open TASEP and the open partially ASEP (PASEP). 
In Section 6, we introduce an alternate joined system
 of the queueing process and the open TASEP. 
Section 7 is devoted to the conclusion of this paper.

\section{review of the m/m/1 queueing process
and the open TASEP}

\subsection{M/M/1 queueing process}

\begin{figure}
\includegraphics{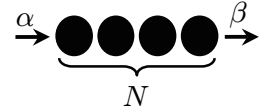}
\caption{\label{fig:queue} M/M/1 queueing process.}
\end{figure}

Let us consider the simplest
 queueing process, i.e., the M/M/1 queueing process,
 as in Fig. \ref{fig:queue}, where $N$ denotes the number of particles.
Particles enter the system  with rate $\alpha$
 and exit the system with rate $\beta$.
(Customers arrive at the queue with rate $\alpha$
 and receive service with rate $\beta$ at one server.)
The M/M/1 queueing process does not have spatial structure and is characterized
 only by the number of particles.

The system is encoded by
 a Markov process on the state space ${\mathbb Z}_{\ge 0}$
 and governed by the following master equation
 for the probability $P(N)$ that the number of particles is $N$:
\begin{align}
 \frac{d}{dt}P(0)=&
 \beta P(1) - \alpha P(0), \\
 \frac{d}{dt}P(N)=&
 \alpha P(N-1) + \beta P(N+1)
  -(\alpha+\beta) P(N) ,
\end{align}
for $N\in \mathbb N$.
The M/M/1 queueing process is equivalent to a continuous-time random walk
 on ${\mathbb Z}_{\ge 0}$ whose jump rates to right and left directions are
 $\alpha$ and $\beta$, respectively, with reflection at 0.

A unique stationary-state solution is easily obtained as
\begin{align}\label{qsol}
 P(N)=& \frac{1}{Z}
 \left(\frac{\alpha}{\beta}\right)^N,
 \quad
 Z = \sum_{N=0}^{\infty}\left(\frac{\alpha}{\beta}\right)^N.
\end{align}
In other words, the equation
\begin{align}\label{rec1}
 0=& \beta P(1) - \alpha P(0), \\
 0=& \alpha P(N-1) + \beta P(N+1) - (\alpha+\beta) P(N)
 \quad (N\in \mathbb N).
 \label{rec2}
\end{align}
with $\sum_{N=0}^{\infty}P(N)=1$ imposed
 has the unique solution \eqref{qsol}.
We should note, however, that
 the normalization constant $Z$ does not always converge,
 in other words, the stationary state does not always exist.
If $\alpha<\beta$, $Z$ actually converges to $\beta/(\beta-\alpha)$
 and the system has the stationary state.
 (Otherwise, $Z$ diverges and the system has no stationary state.)
Thus the critical line is $\alpha=\beta$.
We can see the uniqueness of the stationary state
 recursively as follows:
\begin{itemize}
 \item
   We can set $\displaystyle P(0)=\frac{1}{Z}$.
 \item
   From \eqref{rec1}, we have
   $\displaystyle P(1)=\frac{1}{Z}\frac{\alpha}{\beta}$.
 \item
   The relation \eqref{rec2} implies that
   if we suppose
  \begin{align}
   P(N-1)=\frac{1}{Z}\left(\frac{\alpha}{\beta}\right)^{N-1},\quad
   P(N)=\frac{1}{Z}\left(\frac{\alpha}{\beta}\right)^N
  \end{align}
   for $N\in \mathbb N$,
   then
  \begin{align}
   P(N+1)=\frac{1}{Z}\left(\frac{\alpha}{\beta}\right)^{N+1}.
  \end{align}
\end{itemize}

The average number of particles can be easily calculated as
\begin{align}
 \langle N \rangle_{\rm M/M/1} =\sum_{N=0}^{\infty} NP(N)
 =\frac{\alpha}{\beta-\alpha}.
\end{align}
In the stationary state,
 the current of particles through
 the server is nothing but the arrival rate $\alpha$:
\begin{align}
 \beta\sum_{N=1}^{\infty}P(N)=\alpha.
\end{align}

\subsection{TASEP with open boundaries}

\begin{figure}
\includegraphics{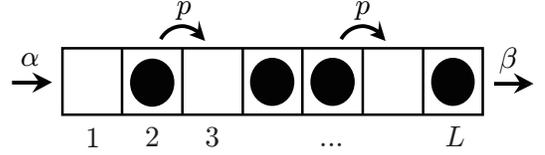}
\caption{\label{fig:tasep} TASEP with open boundaries.}
\end{figure}

Let us consider an interacting particle system,
 the totally asymmetric simple exclusion process
 on the $L$-site chain with open boundaries (open TASEP), see Fig. \ref{fig:tasep}.
Each site can be occupied by
 at most one particle.
Each particle enters the chain
 at the left end with rate $\alpha$ if it is empty,
 hops to its right nearest-neighbor site
 in the bulk with rate $p$ if it is empty,
 and exits at the right end with rate $\beta$.
Let us write $\tau_j=0$ if the $j$th site is empty
and $\tau_j=1$ if it is occupied by a particle.
The system is formulated by
 a Markov process on the state space $\{0,1\}^L$.
The master equation on the probability
 $P(\tau_1,\dots,\tau_L)$
 of finding a configuration $(\tau_1,\dots,\tau_L)$
 is as follows:
\begin{align}\label{tasep_master}
\begin{split}
&   \frac{d}{dt}P(\tau_1,\dots,\tau_L)  \\
  &= \alpha (2\tau_1-1) P(0,\tau_2,\dots,\tau_L) \\
  &\quad+ p\sum_{j=1}^{L-1} (\tau_{j+1}-\tau_j)
     P(\tau_1,\dots,\stackrel{j}{1},
     \stackrel{j+1}{0},\dots,\tau_L) \\
   &\quad+ \beta (1-2\tau_L) P(\tau_1,\dots\tau_{L-1},1).
\end{split}
\end{align}

In contrast to the M/M/1 queueing process,
 the state space of the open TASEP is finite
 and thus it always has a stationary state.
Moreover the open TASEP is irreducible,
 therefore the stationary state is unique.
Derrida et al found
 the stationary-state solution to the open TASEP
 in the following simple form \cite{DEHP}:
\begin{align}\label{mpf}
   P(\tau_1,\dots,\tau_L)=
   \frac{1}{Z_L(\alpha,\beta,p)}
  \langle W({\textstyle \frac{\alpha}{p}})| X_{\tau_1}\cdots X_{\tau_L}
  | V({\textstyle \frac{\beta}{p}}) \rangle 
\end{align}
where $X_0=E$ and $X_1=D$ are matrices,
 $\langle W(u)|$ and $|V(v)\rangle$ are
 row and column vectors, respectively, and
$Z_L(\alpha,\beta,p)$ is the normalization constant
\begin{align}
  Z_L(\alpha,\beta,p)
  = \langle W({\textstyle \frac{\alpha}{p}})| (D+E)^L |V({\textstyle \frac{\beta}{p}})\rangle .
\end{align}
The matrices and the vectors
  should satisfy the following relation
  so that the matrix product form \eqref{mpf}
  actually gives the stationary-state probability:
\begin{align}
\begin{split}
\label{alg}
  DE &= D+E, \\
  \langle W(u)| E & = \frac{1}{u}  \langle W(u)|, \\
  |V(v)\rangle D  & = \frac{1}{v} |V(v)\rangle.
\end{split}
\end{align}
Set $\langle W(u)|V(v)\rangle=1$
 without loss of generality.
The following representation satisfies the algebra \eqref{alg}:
\begin{align}\label{repD}
 D=
  \left(\begin{array}{ccccc}
   1 & 1 & 0 &     0 \\
   0 & 1 & 1 &     0 \\
   0 & 0 & 1 &     1 \\
      &   &  & \ddots &  \ddots     
  \end{array}\right),\ 
   E=
  \left(\begin{array}{ccccc}
   1 & 0 & 0 &  \\
   1 & 1 & 0 &  \\
   0 & 1 & 1 &  \\
      &   &  \ddots & \ddots   
  \end{array}\right),
\\
 \langle W(w)|=\kappa(1,a,a^2,\dots),\ 
 |V(v) \rangle =\kappa
  \left(\begin{array}{c}1\\ b \\ b^2 \\ \vdots
  \end{array}\right)
\label{repWV}
\end{align}
where $\kappa=\sqrt{(w+v-1)/wv}, a=(1-w)/w$ and $b=(1-v)/v$.
Note that, in the original paper \cite{DEHP},
 the bulk hopping rate $p$
 was set to be 1
 and other representations of the matrices and vectors
 were found.
For simplicity in what follows, however,
 we choose the representation \eqref{repD}-\eqref{repWV}
 so that only the vectors depend on $\alpha,\beta$ and $p$.
By using the algebraic relation \eqref{alg},
 we can calculate the normalization constant as follows \cite{DEHP}:
\begin{align}\label{oriform}
 Z_L (\alpha,\beta,p) =
 \sum_{j=0}^{L} \frac{j(2L-j-1)!}{L!(L-j)!}
 \frac{\left(\frac{p}{\alpha}\right)^{j+1}-\left(\frac{p}{\beta}\right)^{j+1}}
 {\frac{p}{\alpha}-\frac{p}{\beta}} .
\end{align}

\begin{figure}[h]
\includegraphics{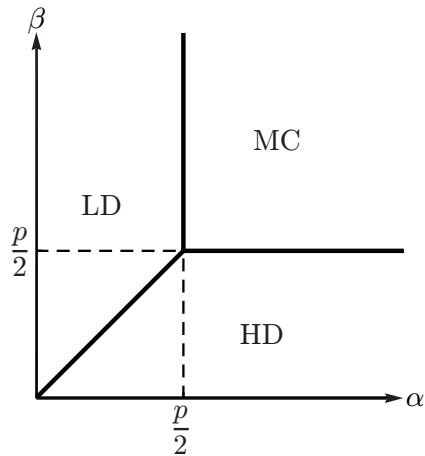}
\caption{\label{fig:tasepa}
Phase diagram of the open TASEP.}
\end{figure}

The stationary current, for example,
 can be written in terms of
 the normalization constant $Z_L(\alpha,\beta,p)$ as
\begin{align}
 J_L(\alpha,\beta,p)=
 p\frac{Z_{L-1}(\alpha,\beta,p)}
 {Z_L(\alpha,\beta,p)}.
\end{align}
In the limit $L\to \infty$,
 the phase diagram of  the current
 $J_{\infty}(\alpha,\beta,p)$
 consists of three regions which are called
 the maximal-current (MC) phase,
 the low-density (LD) phase
 and the high-density (HD) phase
 (see Fig. \ref{fig:tasepa}):
\begin{align}\label{inftycurrent}
 J_{\infty}(\alpha,\beta,p)
 =
 \begin{cases}
  \frac{p}{4}
 & \alpha,\beta\ge\frac{p}{2} \quad \text{(MC)}\\
  \alpha(1-\alpha/p) 
 & \alpha<\min(\beta,\frac{p}{2}) \quad \text{(LD})\\
  \beta(1-\beta/p)
 &  \beta<\min(\alpha,\frac{p}{2})  \quad \text{(HD)}
 \end{cases}
\end{align}
The line $\alpha=\beta<\frac{p}{2}$
 is called the coexistence line,
 where a shock between a low density segment 
 and a high density segment exhibits 
 a random walk \cite{KSKS}:
 $J_{\infty}(\alpha,\alpha,p)=\alpha(1-\alpha/p)$.

Note that no particle can enter the chain
 if the leftmost site is occupied by 
 another particle
 and thus the stationary current is not equal to $\alpha$. 
This means that
 the open TASEP is a ``call-loss system.''
Recall that the stationary current of the M/M/1 queueing process,
 which is not a call-loss system, is $\alpha$.

\section{model}

\begin{figure}
\includegraphics{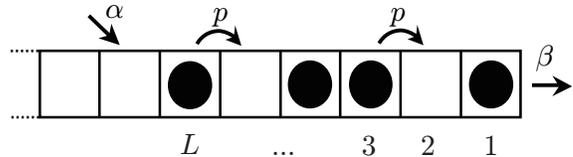}
\caption{\label{fig:model} Queueing process with
 the excluded-volume effect.}
\end{figure}

Let us introduce a new model which is an extended M/M/1 queueing process on a semi-infinite chain
 with the excluded-volume effect as in the TASEP.
Figure \ref{fig:model} shows the model,
 where each site is numbered from right to left and $L$ denotes the leftmost occupied site.
Each site can be occupied by at most one particle.
Each particle hops to its right nearest-neighbor site
 with rate $p$ in the bulk if it is empty and exits with rate $\beta$ at the right end.
The rules of the bulk hopping and the exit are the same as for the open TASEP.
However, the rule for the entering is different.
Each particle enters the chain
 at the immediately left of the leftmost occupied site (or at site 1 if there is no particle on the chain).
One can put this model into the TASEP in the semi-infinite chain with a new boundary condition.

The system is encoded by
 a Markov process on the state space
\begin{align}
\begin{split}
 S&:= \emptyset\cup
 \bigcup_{L\in\mathbb N}
 \Big( \left\{1\right\} \times \{0,1\}^{L-1}
 \Big) \\
 &=
 \{\emptyset,(1),(1,0),(1,1),(1,0,0), \\
  &\quad\quad (1,0,1),(1,1,0),(1,1,1),(1,0,0,0),\dots\}.
\end{split}
\end{align}
The leftmost particle of each element of $S$ except $\emptyset$
 can always be specified.
Configurations such as $(\dots,1,1,1,1)$ and $(\dots,1,0,1,0,1,0)$ do not appear in $S$.

Let $P(1,\tau_{L-1},\dots,\tau_1)$ be
the probability of finding
a configuration $(1,\tau_{L-1},\dots,\tau_1)$
with the leftmost particle at $L$th site
and $P (\emptyset)$
be the probability of finding
no particle on the chain.
The master equation governing the model is as follows:
\begin{align}
&  \frac{d}{dt}P (\emptyset)
  =\beta P (1) - \alpha P (\emptyset) ,\\
&  \frac{d}{dt}P (1)
  = \alpha P(\emptyset) + p P (1,0)
    - (\alpha+\beta) P (1) ,
\end{align}
\begin{align}
\begin{split}
&   \frac{d}{dt}P(1,\tau_{L-1},\dots,\tau_1) \\
  &= pP(1,0,\tau_{L-1},\dots,\tau_1)\\
   &\quad+\alpha\tau_{L-1}
               P(1,\tau_{L-2},\dots,\tau_1)
   - \alpha P(1,\tau_{L-1},\dots,\tau_1) \\
   & \quad+p(\tau_{L-1}-1) P(1,0,\tau_{L-2}\dots,\tau_1) \\
   & \quad+p\sum_{j=1}^{L-2}
    (\tau_j-\tau_{j+1})
    P(1,\tau_{L-1},\dots,\stackrel{j+1}{1},\stackrel{j}{0},
         \dots,\tau_1) \\
    &\quad + \beta (1-2\tau_1)
    P(1,\tau_{L-1},\dots\tau_2,1).
\end{split}
\end{align}
For example,
\begin{align}
\begin{split}
 &\frac{d}{dt}P(1,1,0,1,0,1)\\
&  =\alpha P(1,0,1,0,1) + pP(1,0,1,0,1,0,1) \\
 &\quad+ pP(1,1,1,0,0,1) + pP(1,1,0,1,1,0)\\
  &\quad-  (\alpha+2p+\beta)P(1,1,0,1,0,1), 
\end{split} \\
\begin{split}
 &\frac{d}{dt}P(1,0,0,1,1,0,1,0)\\
 &=pP(1,0,0,0,1,1,0,1,0) + pP(1,0,1,0,1,0,1,0) \\
 &\quad + pP(1,0,0,1,1,1,0,0) + \beta P(1,0,0,1,1,0,1,1)\\
 &\quad - (\alpha+3p)P(1,0,0,1,1,0,1,0).
\end{split}
\end{align}
We write the right-hand side of the master equation
 as $(HP)(\tau_1,\dots,\tau_L)$
 with the generator matrix $H$
 acting on the probability vector
 $P=(P(\emptyset),P(1),P(1,0),\dots)^{\rm T}$.
Note that $H$ is an infinite dimensional matrix.

\section{stationary state}
\subsection{Matrix product form}
The problem is how to find the solution to $HP=0$.
We can see that a slightly arranged matrix product form
\begin{align}
\begin{split}\label{sampf}
  & P(\emptyset)=\frac{1}{\cZ(\alpha,\beta,p)}, \\
  & P(1)=\frac{1}{\cZ(\alpha,\beta,p)} \frac{\alpha}{\beta}
          =\frac{1}{\cZ(\alpha,\beta,p)}\frac{\alpha}{\beta}
          \langle W(1)|V({\textstyle \frac{\beta}{p}})\rangle, \\
  & P(\tau_L=1,\dots,\tau_1) \\
  &\quad=\frac{1}{\cZ(\alpha,\beta,p)}\frac{\alpha^L}{\beta p^{L-1}}
     \langle W(1)| X_{\tau_{L-1}}\cdots
     X_{\tau_1} |V({\textstyle \frac{\beta}{p}})\rangle \\
 & \quad\quad\quad({\rm for}\ L\ge2)
\end{split}
\end{align}
gives a stationary-state solution.
(This idea is applicable to a discrete-time version of the model \cite{AY}.)
Here $\cZ(\alpha,\beta,p)$ is the normalization constant
 which can be written as a special case of the generating function
 of the normalization constant of the open TASEP:
\begin{align}
\begin{split}
   \cZ(\alpha,\beta,p)=&
   \sum_{L=0}^{\infty}\frac{\alpha^L}{\beta p^{L-1}}
   \langle W(1)| (D+E)^{L-1} |V({\textstyle \frac{\beta}{p}})\rangle \\
   =&  \sum_{L=0}^{\infty}\frac{\alpha^L}{\beta p^{L-1}}
           Z_{L-1}(p,\beta,p)
\end{split}
\end{align}
with $Z_{-1}(p,\beta,p)=\beta/p$.
See Appendix \ref{example} for some stationary probabilities
 calculated by using the algebraic relation \eqref{alg}.
That the form \eqref{sampf} gives a stationary-state solution
 (i.e. $HP=0$) can be proved by a similar canceling
 to that for the open TASEP.
Let us use a short-hand notation
\begin{align}
 \langle W(1)| \cdots |({\textstyle \frac{\beta}{p}})\rangle =
 \langle \cdots \rangle.
\end{align}
Substituting the form \eqref{sampf}
 into $HP(1,\tau_{L-1},\dots,\tau_1)$
 and multiplying it by
 $\frac{\beta p^{L-1}}{\alpha^L}\cZ(\alpha,\beta,p)$,
 we find
\begin{align}\label{proof1}
  & \frac{\beta p^{L-1}}{\alpha^L}\cZ(\alpha,\beta,p)
      (HP)(1,\tau_{L-1}\dots,\tau_1) \\
\begin{split}\label{proof2}
=& p\frac{\alpha}{p}
  \langle EX_{\tau_{L-1}}\cdots X_{\tau_1}\rangle \\
  & +\alpha\tau_{L-1}\frac{p}{\alpha}
    \langle X_{\tau_{L-2}} \cdots X_{\tau_1}\rangle
 -\alpha\langle X_{\tau_{L-1}}\cdots X_{\tau_1}\rangle \\
 & + p(\tau_{L-1}-1)
  \langle EX_{\tau_{L-2}}\cdots X_{\tau_1}\rangle \\
 & + p\sum_{j=1}^{L-2}(\tau_j-\tau_{j+1})
  \langle X_{\tau_{L-1}}
    \cdots \stackrel{j+1}{D}
    \stackrel{j}{E} \cdots X_{\tau_1}\rangle \\
 & + \beta(1-2\tau_1)
  \langle X_{\tau_{L-1}}\cdots X_{\tau_2}D\rangle
\end{split} \\
\begin{split}\label{proof3}
=& p\big\{
       -(1-2\tau_{L-1})
       \langle X_{\tau_{L-2}}\cdots X_{\tau_1}\rangle \\
&+ \sum_{j=1}^{L-2} (1-2\tau_{j+1})
       \langle X_{\tau_{L-1}}\cdots X_{\tau_{j+2}}X_{\tau_j} \cdots X_{\tau_1}\rangle \\
&- \sum_{j=1}^{L-2} (1-2\tau_j)
       \langle X_{\tau_{L-1}}\cdots X_{\tau_{j+1}}X_{\tau_{j-1}} \cdots X_{\tau_1}\rangle\\
&+  (1-2\tau_1)
  \langle X_{\tau_{L-1}}\cdots X_{\tau_2}\rangle\big\}
\end{split} \\
=& 0 \label{proof4}
\end{align}
In going from \eqref{proof2} to \eqref{proof3},
 we applied the algebraic relation \eqref{alg}:
\begin{align}
 \langle\cdots DE\cdots\rangle
 =&\langle\cdots D\cdots\rangle
   + \langle\cdots E\cdots\rangle,\\
 \langle E\cdots \rangle
 =&\langle \cdots\rangle, \quad
 \beta\langle\cdots D\rangle
 = p\langle\cdots \rangle.
\end{align}

In our argument throughout this section, we assume
 that \eqref{sampf} is a unique stationary state of the model
 if $\cZ(\alpha,\beta,p)$ converges and there is no stationary state 
 if $\cZ(\alpha,\beta,p)$ diverges.
Recall that the stationary states of the usual M/M/1 queueing process
 and the open TASEP are unique.

The following function will be useful in the next section:
\begin{align}
\begin{split}
&
 \cZ(p\xi,\beta,p;\zeta) \\
=& 1+
   \sum_{L=1}^{\infty}\frac{p}{\beta}\xi^L \zeta
   \langle W(1)| (\zeta D+E)^{L-1} |V({\textstyle \frac{\beta}{p}})\rangle
\end{split}
\end{align}
with a fugacity $\zeta$.
This can be regarded as a special case of the generating function
 of the normalization constant of the TASEP with a single defect particle
 (see section 4.3 in \cite{BE}). 
The case where $\xi=\alpha/p$ and $\zeta=1$
 corresponds to the normalization constant:
\begin{align}
\begin{split}
 \cZ(\alpha,\beta,p) 
= \cZ(p\, \alpha/p,\beta,p;1) .
\end{split}
\end{align}

\subsection{Critical line}
\begin{figure}
\includegraphics{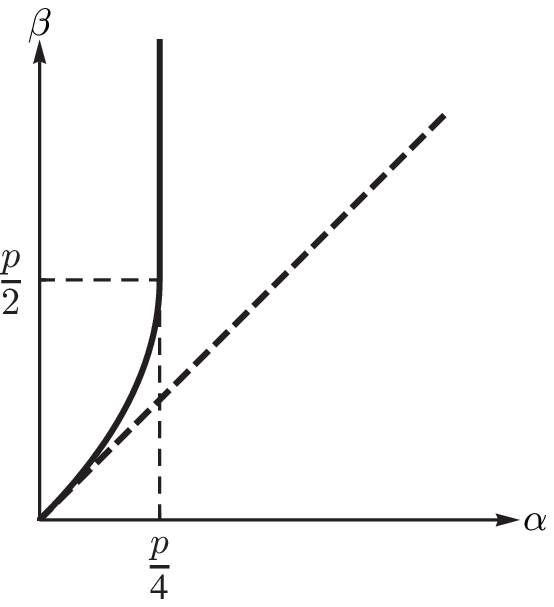}
\caption{\label{fig:cline} Critical line \eqref{cl}.}
\end{figure}

The asymptotic behavior of $Z_L(p,\beta,p)$
 as $L\to \infty$ is as follows \cite{DEHP}:
\begin{align}
Z_L(p,\beta,p)
\sim
\begin{cases}
\frac{4^L}{\sqrt{\pi L^3}}
\left(\frac{2\beta}{2\beta-p}\right)^2
& \beta>\frac{p}{2} \\
\frac{2\cdot 4^L}{\sqrt{\pi L}}
& \beta=\frac{p}{2} \\
\frac{1-2\beta/p}
 {(1-\beta/p)^2}
\left(\frac{1}{(1-\beta/p)\beta/p}\right)^L
 & \beta<\frac{p}{2} 
\end{cases}.
\end{align}
Thanks to this asymptotic form, we see that if the condition
\begin{align}\label{cond}
 \begin{cases}
   \alpha\le\frac{p}{4}  & (\beta > \frac{p}{2})\\
  \alpha<\beta(1-\beta/p)  & (\beta\le\frac{p}{2})
 \end{cases}
\end{align}
is satisfied, then $\cZ(\alpha,\beta,p)$ converges.
In other words, the critical line is given by
\begin{align}\label{cl}
 \alpha=\alpha_c=
 \begin{cases}
  \frac{p}{4} & (\beta>\frac{p}{2})\\
  \beta(1-\beta/p) & (\beta\le\frac{p}{2})
 \end{cases},
\end{align}
see Fig. \ref{fig:cline}.
Note that the area \eqref{cond},
 where the normalization constant $\cZ(\alpha,\beta,p)$ converges,
 is embedded in that of the usual M/M/1 queueing process ($\alpha<\beta$).
We remark that the critical line can be written in terms
 of the stationary current \eqref{inftycurrent} of the open TASEP with $\alpha=p$:
\begin{align}
 \alpha_c=J_{\infty}(p,\beta,p).
\end{align}

Turning to the stationary current through
 the right end of the chain,
 we see that this must be the arrival rate $\alpha$,
 because the model is not a call-loss system.
In fact, one can see
\begin{align}
\begin{split}
 & \beta P(1)+ \beta \sum_{L=2}^{\infty}\sum_{\tau_j=0,1}
 P(1,\tau_{L-1},\dots,\tau_2,1) 
 \\
 =& \frac{\alpha}{\cZ(\alpha,\beta,p)} \\
 &\quad
  +\frac{1}{\cZ(\alpha,\beta,p)}\sum_{L=2}^{\infty}
  \frac{\alpha^L}{p^{L-1}}
   \langle W(1)| (D+E)^{L-2} D|V({\textstyle \frac{\beta}{p}})\rangle
 \\
 =& \frac{\alpha}{\cZ(\alpha,\beta,p)}\left(1+
  \sum_{L=2}^{\infty}\frac{\alpha^{L-1}}{\beta p^{L-2}}
   \langle W(1)| (D+E)^{L-2}|V({\textstyle \frac{\beta}{p}})\rangle
    \right)
 \\
 =& \alpha.
\end{split}
\end{align}

\subsection{Average values}
In this subsection and the next subsection, we assume
 that the condition \eqref{cond} is satisfied.
According to the formula (4.27) in \cite{BE}
 we find
\begin{align}
\begin{split}
& \cZ(p\xi,\beta,p;\zeta) \\
&=1+\frac{p\xi\zeta}{\beta\left(1-\xi(\sqrt\zeta-1)^2 \right)
       \left(1- \eta\right)
       \left(1- \eta\mu \right)
},
\end{split}
\end{align}
where
\begin{align}
\mu &=1+\sqrt{\zeta}\left(p/\beta-1\right),\\
\eta &=\frac{1}{2}\left(1-\sqrt{\frac{1-(1+\sqrt{\zeta})^2\xi}{1-(1-\sqrt
{\zeta})^2\xi}}\right)\end{align}
In particular,
\begin{align}
 \cZ(\alpha,\beta,p)
 =\frac{2\beta}{2\beta-p(1-r)}
\end{align}
with $r=\sqrt{1-4\alpha/p}$.
The average length $\langle L \rangle$ (the average position of the leftmost particle) 
 and the average number  $\langle N \rangle$
 of particles on the chain can be calculated
 by differentiating $\cZ(p\xi,\beta,p;\zeta)$ as
\begin{align}\label{<L>}
\langle L \rangle =&
   \xi \frac{\partial}{\partial\xi}
    \ln \cZ(p\xi,\beta,p;1)
    \Big|_{\xi=\alpha/p} 
    = \frac{2\alpha/p}{r(-1+r+2\beta/p)},
\\
\langle N \rangle =&
   \frac{\partial}{\partial\zeta}
    \ln \cZ(\alpha,\beta,p;\zeta)
    \Big|_{\zeta=1} 
  = \frac{2\alpha(1+r-(3+r)\alpha/p)}
         {r(1+r)((1+r)\beta-2\alpha)}.
 \label{<N>}
\end{align}
By the excluded-volume effect,
 these values are greater than the average length of
 the usual M/M/1 queueing process:
\begin{align}\label{grgreq}
 \langle L\rangle >
 \langle N\rangle >
 \langle N\rangle_{\rm M/M/1}
 =\frac{\alpha}{\beta-\alpha}.
\end{align}
Actually, one can find
\begin{align}
 \langle N\rangle -
 \langle N\rangle_{\rm M/M/1}
 = \frac
  {2\alpha^2\left((\beta-\alpha)+r(\alpha+\beta)\right)}
  {pr(1+r)(\beta(1+r)-2\alpha)(\beta-\alpha)}
 >0,
\end{align}
whereas the first inequality
 in \eqref{grgreq} is true by definition.
$\langle L\rangle$ and
 $\langle N\rangle$
 are expanded with respect to $1/p$ as
\begin{align}
 \langle L \rangle
    \sim& \frac{\alpha}{\beta-\alpha}
      +\frac{\alpha^2(2\beta-\alpha)}
               {(\beta-\alpha)^2}
          \frac{1}{p}
        + O\left((1/p)^2\right), \\
 \langle N \rangle
    \sim& \frac{\alpha}{\beta-\alpha}
      +\frac{\alpha^2\beta}
               {(\beta-\alpha)^2}
           \frac{1}{p}
        + O\left((1/p)^2\right)
\end{align}
and we can see a natural result
 that they approach
 $\langle N\rangle_{\rm M/M/1}$
 in the usual-M/M/1-queuing-process
 limit $p\to\infty$.
Note that
\begin{align}
 \langle L\rangle = \infty,\quad
 \langle N\rangle = \infty
\end{align}
on the critical line $\alpha=\frac{p}{4}$
 and $\beta>\frac{p}{2}$.

\subsection{Monotonicity of the length}

Let us consider
 the probability $\lambda_L$ that the length is $L$
 (the leftmost particle is at site $L$):
\begin{align}
\begin{split}
 \lambda_L =&
 \frac{1}{\cZ(\alpha,\beta,p)}
      \sum_{\tau_i=0,1} P(1,\tau_{L-1},\dots,\tau_1) \\
 =& \frac{1}{\cZ(\alpha,\beta,p)}
        \frac{p}{\beta}
        \left(\frac{\alpha}{p}\right)^L
        Z_{L-1}(p,\beta,p)
\end{split}
\end{align}
For $L=0$, we set $\lambda_0=P(\emptyset)$.

Thanks to the asymptotic form again \eqref{sampf},
 we can see that $\lambda_L$ decays as $L\to\infty$ as
\begin{align}
\lambda_L
\sim
\frac{1}{\cZ(\alpha,\beta,p)}
        \frac{p}{\beta}
 \times
\begin{cases}
\frac{1}{\sqrt{\pi L^3}}
\left(\frac{2\beta}{2\beta-p}\right)^2
 \left(\frac{4\alpha}{p}\right)^L
& \beta>\frac{p}{2} \\
\frac{2}{\sqrt{\pi L}}\left(\frac{4\alpha}{p}\right)^L
& \beta=\frac{p}{2} \\
\frac{1-2\beta/p}
 {(1-\beta/p)^2}
\left(\frac{\alpha}{(1-\beta/p)\beta}\right)^L
 & \beta<\frac{p}{2} 
\end{cases}.
\end{align}

When $L$ is finite, $\lambda_L$ possesses the property of
 the monotonicity with respect to $L$:
\begin{align}\label{mono}
 \cdots<\lambda_2<\lambda_1<\lambda_0.
\end{align}
The rightmost inequality $\lambda_1<\lambda_0$ is clearly true.
We devote the rest of this subsection
 to the proof of $\lambda_{L+1}<\lambda_L$ for $L\ge1$. 
Let us use short-hand notations
\begin{align}
 Z_L:=Z_L(p,\beta,p),\quad
 x:=\frac{p}{\beta},
\end{align}
and the following alternate expression
which can be obtained by transforming \eqref{oriform}:
\begin{align}
 Z_L=\sum_{j=0}^{L}a_{L,j}\,x^j,
\end{align}
where $a_{L,j}=\frac{(j+1)(2L-j)!}{(L+1)!(L-j)!}$.

\subsubsection{Case when $\beta\le p/2$}
Under the assumption
 $\alpha<\alpha_c=p/x(1-1/x)$,
 we find that
\begin{align}
 &\frac{\alpha}{p} Z_{L} - Z_{L-1}
 < \frac{1}{x}\left(1-\frac{1}{x}\right)
 Z_{L} - Z_{L-1} 
  = - \frac{ C_L }{x^2}
 <0,
\end{align}
where $C_L=\frac{(2L)!}{(L+1)!L!}$
 is the Catalan number.
Thus, we have $\lambda_{L+1}<\lambda_L$.

\subsubsection{Case when $\beta> p/2$}
The proof of the monotonicity for $\beta > p/2$
 (i.e. $x< 2$)
 will be somewhat more technical.
Our goal is to show $Z_L<4Z_{L-1}$, this implies that $\frac{\alpha}{p}Z_L\le\frac{1}{4}Z_L<Z_{L-1}$
 and thus $\lambda_{L+1}<\lambda_L$.

Before proving this in the general $L\ge 1$ case,
 we demonstrate it for $L=6$:
\begin{align}
 \begin{split}
 \label{L=6}
   Z_6
      &=132+132x+90x^2+48x^3+20x^4+6x^5+x^6 \\
      &<132+132x+90x^2+48x^3+20x^4+8x^5 \\
      &<132+132x+90x^2+48x^3+28x^4+4x^5 \\
      &<132+132x+90x^2+64x^3+20x^4+4x^5 \\
      &<132+132x+106x^2+56x^3+20x^4+4x^5 \\
      &<168+168x+112x^2+56x^3+20x^4+4x^5 \\
  &= 4Z_5,
\end{split}
\end{align}
where we used $x^6<2x^5$, $x^5<2x^4$, etc.

Let us go back to the general $L\ge 1$ case and
 introduce a sequence $\{b_{L,j}\}_{1\le j\le L-1}$
 defined by the following recursion relation:
\begin{align}
 b_{L,j}=& 2(b_{L,j+1} - 4a_{L-1,j+1}) + a_{L,j}
\end{align}
with $a_{L-1,L}=b_{L,L+1}=0$.
One can find that
\begin{align}
\begin{split}
 b_{L,j}= &
 \frac{(2L-j-2)!}{(L+1)!(L-j)!} \\
  &\quad \times (4jL^2-2(j^2-4j-3)L-j(j+1)(j+7)).
\end{split}
\end{align}
As long as $b_{L,j}>4a_{L-1,j}$,
\begin{align}
\begin{split}\label{inequality}
 &  a_{L,j-1}x^{j-1} + b_{L,j}x^j \\
 =& a_{L,j-1}x^{j-1} + (b_{L,j}-4a_{L-1,j})x^j + 4a_{L-1,j}x^j  \\
 <& a_{L,j-1}x^{j-1} + 2(b_{L,j}-4a_{L-1,j})x^{j-1} + 4a_{L-1,j}x^j  \\
 =& b_{L,j-1}x^{j-1} + 4a_{L-1,j}x^j .
\end{split}
\end{align}
Let $k$ be
 an integer such that
\begin{align}
 b_{L,k}\le 4a_{L-1,k} \text{\quad and \quad}
 b_{L,j}>4a_{L-1,j}  \ (\forall j>k).
\end{align}
This is equivalent to
\begin{align}\label{kk}
  k(k+3) \le 2L < (k+1)(k+4)
\end{align}
and $k$ is determined uniquely.
For example, $k=2$ for $L=6$.
Using the inequality \eqref{inequality} repeatedly while $j>k$, we get
\begin{align}
\begin{split}
\label{mochotto}
Z_L &= a_{L,0}
        +\cdots+a_{L,L-2}x^{L-2}+a_{L,L-1}x^{L-1}+a_{L,L}x^L \\
  &< a_{L,0}
        +\cdots+a_{L,L-2}x^{L-2}+b_{L,L-1}x^{L-1} \\
   &< a_{L,0}
        +\cdots+b_{L,L-2}x^{L-2}+4a_{L-1,L-1}x^{L-1}  \\
   &< \cdots \\
   &< \sum_{j=0}^{k-1} a_{L,j}x^j + b_{L,k} x^{k}
         + \sum_{j=k+1}^{L-1} 4a_{L-1,j}x^j .
\end{split}
\end{align}
The coefficients $\{a_{L,j}\}_{0\le j\le k-1}$
 in the first summation of the last line of \eqref{mochotto}
 satisfy $a_{L,j}<4a_{L-1,j}$:
\begin{align}
\begin{split}
\because\quad \quad 
   & a_{L,j} - 4a_{L-1,j} \\
 =& \frac{(j+1)(2L-j-2)!(j^2+5j-6L)}{(L+1)!(L-j)!} \\
 <&  \frac{(j+1)(2L-j-2)!(2j-4L)}{(L+1)!(L-j)!} <0,
\end{split}
\end{align}
where we used $j^2+3j<k^2+3k\le 2L$
 (see \eqref{kk}).

Finally, we achieve
\begin{align}
\begin{split}
 Z_L <& \sum_{j=0}^{k-1} 4a_{L-1,j}x^j + 4a_{L-1,k} x^{k}
         + \sum_{j=k+1}^{L-1} 4a_{L-1,j}x^j \\
        =& 4Z_{L-1} .
\end{split}
\end{align}

\section{generalization}

\begin{figure}
\includegraphics{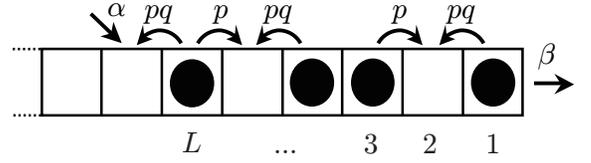}
\caption{\label{fig:qpasep}Queueing process with exclusive hopping
to both directions.}
\end{figure}

\begin{figure}
\includegraphics{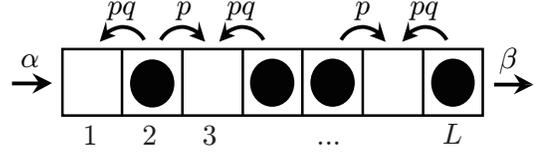}
\caption{\label{fig:pasep}
PASEP with open boundaries.}
\end{figure}

In this section, we generalize the model by allowing particles to hop backward
 with rate $pq\ (q>0)$, see Fig. \ref{fig:qpasep}.
We will construct a stationary state and 
 derive the critical line,
 arranging the matrix product form 
 of the partially ASEP (PASEP)
 with the open boundary condition as in Fig. \ref{fig:pasep}.

The matrix product
  stationary state of the open PASEP
  is as follows:
\begin{align}\label{mpfpasep}
\begin{split}
   &P(\tau_1,\dots,\tau_L) \\
   &=
   \frac{1}{Z_L(\alpha,\beta,p,q)}
  \langle W({\textstyle \frac{\alpha}{p}})| X_{\tau_1}\cdots X_{\tau_L}
  | V({\textstyle \frac{\beta}{p}}) \rangle ,
\end{split}
\end{align}
where the matrices $X_0=E_q$ and $X_1=D_q$,
 the row vector $\langle W(w)|$  and the column vector $|V(v)\rangle$ satisfy
\begin{align}
\begin{split}
\label{alg2}
  D_qE_q-qE_qD_q   &= D_q+E_q, \\
  \langle W(w)| E_q & = \frac{1}{w}  \langle W(w)|, \\
  D_q|V(v)\rangle     & = \frac{1}{v} |V(v)\rangle,
\end{split}
\end{align}
and $Z_L(\alpha,\beta,p,q)$ is the normalization constant:
\begin{align}
  Z_L(\alpha,\beta,p,q)=
  \langle W({\textstyle \frac{\alpha}{p}})| (D_q+E_q)^L
  | V({\textstyle \frac{\beta}{p}}) \rangle  .
\end{align}
Some representations of the matrices and the vectors can be found
 in \cite{BE,BECE,Sa}.
The notation used in \eqref{alg2} was chosen for convenience.
However, we stress that, despite appearances,
 we have found no representation such that
 the matrices depend only on $q$ and the vectors are independent of $q$.

For $q<1$, the normalization constant $Z_L(\alpha,\beta,p,q)$
 can be written in the following integral form \cite{BECE,Sa}:
\begin{align}\label{intform}
\begin{split}
 & Z_L(\alpha,\beta,p,q) \\
 =& \frac{(q,ab;q)_\infty}{4\pi i} 
        \int_K\frac{dz}{z}\frac{(z^2,z^{-2};q)_\infty}
        {(az,a/z,bz,b/z;q)_\infty}
        \left(\frac{2+z+z^{-1}}{1-q}\right)^L
\end{split}
\end{align}
where $a=\frac{p(1-q)}{\alpha}-1,b=\frac{p(1-q)}{\alpha}-1$
 and $(x_1,\dots,x_n;q)_\infty=\prod_{1\le n\le m}\prod_{0\le i\le \infty}(1-x_n q^i)$
 is the $q$-shifted factorial.
The contour $K$ encloses poles $z=a,qa,q^2a,\dots$
 and $z=b,qb,q^2b,\dots$,
 and excludes $z=1/a,q/a,q^2/a,\dots$
 and $z=1/b,q/b,q^2/b,\dots$.
The asymptotic form of the normalizing constant of the open PASEP in the limit $L\to\infty$
 has been obtained by applying the saddle point method to the integral form
 \eqref{intform}  \cite{BECE,Sa}:
\begin{align}\label{forward}
\begin{split}
& Z_L(\alpha,\beta,p,q) \sim \\
&
 \begin{cases}
 \frac{4(ab;q)_{\infty}(q;q)^3_{\infty}}{\sqrt{\pi}(a,b;q)^2_{\infty}L^{\frac{3}{2}}}
        \left(\frac{4}{1-q}\right)^L
  & \alpha,\beta>\frac{p(1-q)}{2}
\\ 
 \frac{2}{\sqrt{\pi}(b;q)_\infty L^{\frac{1}{2}}}\left(\frac{4}{1-q}\right)^L
  &  \alpha=\frac{p(1-q)}{2}<\beta
\\
 \frac{2}{\sqrt{\pi}(a;q)_\infty L^{\frac{1}{2}}}\left(\frac{4}{1-q}\right)^L
  &  \beta=\frac{p(1-q)}{2}<\alpha
\\
 \left(\frac{4}{1-q} \right)^L &  \alpha=\beta=\frac{p(1-q)}{2}
\\
 \frac{(a^{-2};q)_{\infty}}{(b/a;q)_{\infty}}\left(\frac{2+a+a^{-1}}{1-q} \right)^L
  &  \alpha<\min(\beta,\frac{p(1-q)}{2})
\\
 \frac{(b^{-2};q)_{\infty}}{(a/b;q)_{\infty}}\left(\frac{2+b+b^{-1}}{1-q} \right)^L
  & \beta<\min(\alpha,\frac{p(1-q)}{2}) 
\\
 \frac{(a-1)(a^{-2};q)_{\infty}L}{(a+1)(q;q)_{\infty}}
 \left(\frac{2+a+a^{-1}}{1-q} \right)^L
  & \alpha=\beta<\frac{p(1-q)}{2}
\end{cases} .
\end{split}
\end{align}
The stationary current of the open PASEP
 can be written in terms of the normalization constant as
\begin{align}
  J_L(\alpha,\beta,p,q)=
  p\frac{Z_{L-1}(\alpha,\beta,p,q)}{Z_L(\alpha,\beta,p,q)} .
\end{align}
Noting the asymptotic form \eqref{forward}, we have
\begin{align}
 J_\infty(\alpha,\beta,p,q) =
 \begin{cases}
 \frac{p(1-q)}{4} & \alpha,\beta\ge\frac{p(1-q)}{2}\\ 
 \alpha\left(1-\frac{\alpha}{p(1-q)}\right)
   & \beta\le\min(\alpha,\frac{p(1-q)}{2})  \\
 \beta\left(1-\frac{\beta}{p(1-q)}\right)
   & \alpha\le\min(\beta,\frac{p(1-q)}{2}) 
 \end{cases}.
\end{align}

In the symmetric case $q=1$,
 the normalization constant has
 the following simple form \cite{SMW}:
\begin{align}\label{symmetric}
 Z_L(\alpha,\beta,p,q)
   =(\gamma+1)(\gamma+2)\cdots(\gamma+L)
\end{align}
where $\gamma=\frac{p}{\alpha}+\frac{p}{\beta}-1$.

In the reverse-bias case $q>1$,
 the normalization constant behaves
 in the limit $L\to\infty$ as
\begin{align}\label{reverse}
 Z_L(\alpha,\beta,p,q)
 \sim A q^{\frac{L^2}{4}}\left(\frac{\sqrt{ab}}{1-q}\right)^L ,
\end{align}
where $A$ is a constant independent of $L$ \cite{BECE}.

Let us go back to the generalized queueing process
 with forward and backward hopping (Fig. \ref{fig:qpasep}),
 which is governed by the following master equation:
\begin{align}
&  \frac{d}{dt}P (\emptyset)
  =\beta P (1) - \alpha P (\emptyset) ,\\
&  \frac{d}{dt}P (1)
  = \alpha P(\emptyset) + p P (1,0)
    - (\alpha+\beta+pq) P (1) ,\\
\begin{split}
&   \frac{d}{dt}P(1,\tau_{L-1},\dots,\tau_1) \\
  &= pP(1,0,\tau_{L-1},\dots,\tau_1) - pqP(1,\tau_{L-1},\dots,\tau_1)\\
   &\quad+\alpha\tau_{L-1}
               P(1,\tau_{L-2},\dots,\tau_1)
   - \alpha P(1,\tau_{L-1},\dots,\tau_1) \\
   & \quad+p(\tau_{L-1}-1) P(1,0,\tau_{L-2}\dots,\tau_1) \\
   & \quad+pq(1-\tau_{L-1}) P(1,\tau_{L-2}\dots,\tau_1) \\
   & \quad+p\sum_{j=1}^{L-2} (\tau_j-\tau_{j+1})
    P(1,\tau_{L-1},\dots,\stackrel{j+1}{1},\stackrel{j}{0},
         \dots,\tau_1) \\
   & \quad+pq\sum_{j=1}^{L-2} (\tau_{j+1}-\tau_j)
    P(1,\tau_{L-1},\dots,\stackrel{j+1}{0},\stackrel{j}{1},
         \dots,\tau_1) \\
    &\quad + \beta (1-2\tau_1)
    P(1,\tau_{L-1},\dots\tau_2,1).
\end{split}
\end{align}
A stationary-state solution to this equation is given by the following form,
 which can be proved in the same way as \eqref{proof1}--\eqref{proof4}:
\begin{align}
\begin{split}
  & P(\emptyset)=\frac{1}{\cZ(\alpha,\beta,p,q)}, \quad
    P(1)=\frac{1}{\cZ(\alpha,\beta,p,q)} \frac{\alpha}{\beta} ,\\
  & P(1,\tau_{L-1}\dots,\tau_1) \\
  &\ =\frac{1}{\cZ(\alpha,\beta,p,q)}\frac{\alpha^L}{\beta p^{L-1}}
                \langle W({\textstyle \frac{\alpha}{\alpha+pq}} )|
                 X_{\tau_{L-1}}\cdots X_{\tau_1}
                |V({\textstyle \frac{\beta}{p}})\rangle ,
\end{split}
\end{align}
where $\cZ(\alpha,\beta,p,q)$
 is the normalization constant:
\begin{align}
 \cZ (\alpha,\beta,p,q)
  = \sum_{L=0}^{\infty} \frac{\alpha^L}{\beta p^{L-1}}
 Z_{L-1}({\textstyle \frac{\alpha p}{\alpha+pq}},\beta,p,q)
\end{align}
with $Z_{-1}(\frac{\alpha p}{\alpha+pq},\beta,p,q)=\beta/p$.

We now obtain the critical line assuming the uniqueness of the stationary state.
In view of \eqref{symmetric} and \eqref{reverse},
 we find that the normalization constant $\cZ(\alpha,\beta,p,q)$ converges
 only if the hopping ratio $q<1$.
Moreover, using the asymptotic form \eqref{forward}, we find
 that the condition for the model to have the stationary state is
\\
for $0<q<\frac{1}{3}$,
\begin{align}\label{qpasepcl1}
  \begin{cases}
  \alpha\le\alpha_c=\frac{p(1-q)}{4} & \beta>\frac{p(1-q)}{2} \\
  \alpha<\alpha_c=\beta\left(1-\frac{\beta}{p(1-q)}\right)
      & \beta\le\frac{p(1-q)}{2}
 \end{cases},
\end{align}
for $\frac{1}{3}\le q<1$,
\begin{align}\label{qpasepcl2}
 \alpha<\alpha_c=
 \begin{cases}
   \frac{p\left(s-q(3-2q)\right)}{2(1-q)}
        & \beta>\frac{p\left(s+2-q\right)}{2} \\
   \beta\left(1-\frac{\beta}{p(1-q)}\right)
        & \beta\le\frac{p\left(s+2-q\right)}{2}
 \end{cases},
\end{align}
where   $s=\sqrt{q(4-3q)}$.
Note that the critical line $\alpha_c$ is just the solution to
\begin{align}
 \alpha_c=J_\infty({\textstyle\frac{\alpha_c p}{\alpha_c +pq}},\beta,p,q).
\end{align}

\begin{figure}[h]
\includegraphics{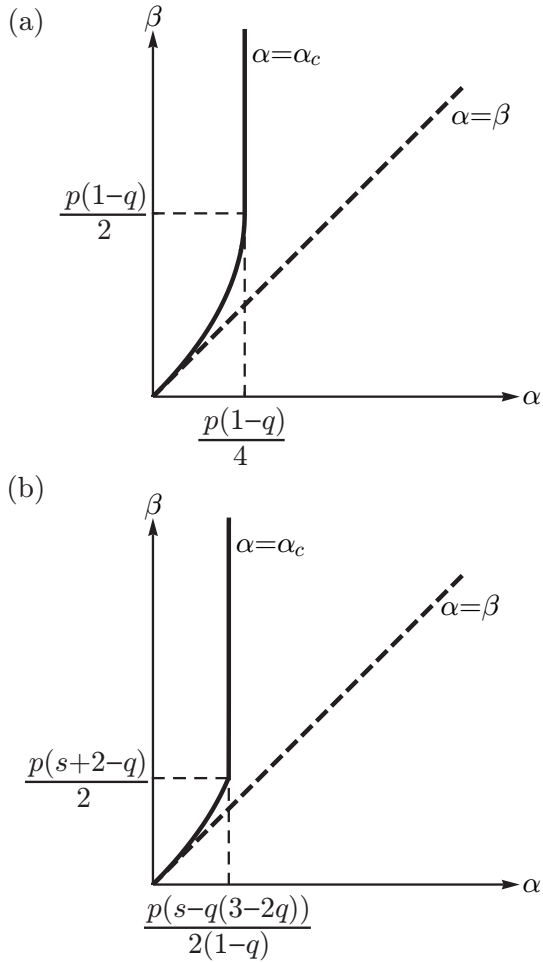}
\caption{\label{fig:pdqpasep}
 Critical lines for
 (a) $0<q<\frac{1}{3}$ and
 (b) $\frac{1}{3}\le q <1$.}
\end{figure}

\section{alternate model}

\begin{figure}[h]
\includegraphics{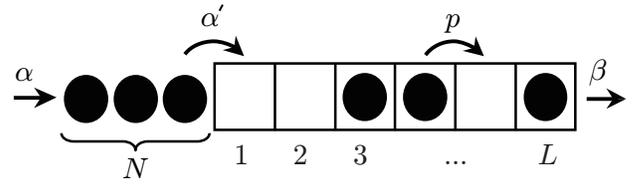}
\caption{\label{fig:alt}
 Alternate joined system of the M/M/1 queueing process and the open TASEP.}
\end{figure}

We introduce here an alternate joined system
 of the M/M/1 queueing process and the open TASEP.
This new system consists of a queue part and a TASEP part, see Fig. \ref{fig:alt}.
Each particle enters the system with rate $\alpha$  and joins the queue part.
The queue part has no spatial structure,
 and is characterized by the number of particles $N$.
Each particle leaves the queue part
 and enters the TASEP part with rate $\alpha'$.
After entering the TASEP part,
 particles follow the same rule as in the usual open TASEP.
This is a model of, for example, a production line
 with a material inventory.

The state space of the Markov process encoding the model 
 is $\mathbb Z _{\ge0}\times \{0,1\}^L$.
The master equation governing the probability $P(N,\tau_1,\dots,\tau_L)$ of
 finding the configuration $(N,\tau_1,\dots,\tau_L)$ is
\\
 \quad
\\
 \quad
\\
 \quad
\\
\begin{align}
\begin{split}
&   \frac{d}{dt}P(N,\tau_1,\dots,\tau_L)  \\
  &= \alpha(1-\delta_{N0})P(N-1,\tau_1,\dots,\tau_L) \\
    &\quad -\alpha P(N,\tau_1,\dots,\tau_L)\\
   &\quad +\alpha'\tau_1P(N+1,0,\tau_2,\dots,1,0,\dots,\tau_L) \\
   &\quad   -\alpha'(1-\tau_1)P(N,0,\tau_2,\dots,1,0,\dots,\tau_L) \\
   &\quad+ p\sum_{j=1}^{L-1} (\tau_{j+1}-\tau_j)
     P(N,\tau_1,\dots,\tau_{j-1}1,0,\tau_{j+2},\dots,\tau_L) \\
   &\quad+ \beta (1-2\tau_L) P(N,\tau_1,\dots\tau_{L-1},1).
\end{split}
\end{align}
For example, with $L=4$,
\begin{align}
\begin{split}
  \frac{d}{dt}P(0,1,1,0,1)
  =&\alpha' P(1,0,1,0,1) + pP(0,1,1,1,0)  \\
       &- (\alpha+p+\beta)P(0,1,1,0,1),
\end{split} \\
\begin{split}
  \frac{d}{dt}P(5,0,1,0,0)
  =&\alpha P(4,0,1,0,0) + pP(5,1,0,0,0) \\
     &+ \beta P(5,0,1,0,1) \\
     &-(\alpha+\alpha'+p)P(5,0,1,0,0).
\end{split}
\end{align}

It is difficult to find
 an exact stationary state of this model.
One can expect, however, the critical line
 separating the parameter space to take the form
\begin{align}\label{alt_a=J}
 \alpha=J_L(\alpha',\beta,p),
\end{align}
because
\begin{itemize}
 \item
 In a stationary state, the current must be $\alpha$.
 \item
 If there is no stationary state and the queue part continues to grow,
 the TASEP part can be regarded as the open TASEP
 with a particle reservoir in the left end.
 Thus, the current must be $J_L(\alpha',\beta,p)$ in this case. 
 \item
 These two values should be equal on the critical line.
\end{itemize}
In fact, the critical lines for $L=1,2,3,4$
 with $\alpha'=p$ were calculated
 (although not rigorously)
 and found to agree with \eqref{alt_a=J} \cite{A}.

\section{Conclusion}

We have studied an extension of the M/M/1 queueing process
 on a semi-infinite chain with the excluded-volume effect as in the open TASEP.
We found that a stationary-state solution is given
 by a slightly arranged matrix product form of the open TASEP
 and its normalization constant is given
 by the generating function of that of the open TASEP.
The critical line which separates the parameter space into the regions
 with and without the stationary state
 is written in terms of the stationary current of the open TASEP.
We also calculated the average length of the system and the average number of particles.
We also showed the monotonicity of the probability of the length.
These were derived by assuming the uniqueness of the stationary state.
We generalized the model by allowing particles to hop backward
 and obtained its critical line.
An alternate joined system of the queueing process and the open TASEP
 was introduced.
We expect that its critical line can be also written in terms of the 
 current of the open TASEP.

\begin{acknowledgments}
The author thanks R. Jiang, R. Nishi, K. Nishinari, S. Saito and T. Shirai
 for fruitful discussion.
He is also grateful to M. Hay for his critical reading of the manuscript.
This work is supported
 by Global COE Program ``Education and Research Hub for Mathematics-for-Industry.''
\end{acknowledgments}

\appendix
\section{example\label{example}}
The stationary probabilities \eqref{sampf} for some configurations
 are listed here:
\begin{align*}
&  \cZ P (1,0)=\frac{\alpha ^2}{p \beta },
\ \cZ P (1,1)=\frac{\alpha ^2}{\beta ^2},
\  \cZ P (1,0,0)=\frac{\alpha ^3}{p^2 \beta },
\\ & \cZ P (1,0,1)=\frac{\alpha ^3}{p \beta ^2},
\ \cZ P (1,1,0)=\frac{\alpha ^3 (p+\beta )}{p^2 \beta ^2},
\\ & \cZ P (1,1,1)=\frac{\alpha ^3}{\beta ^3},
\ \cZ P (1,0,0,0)=\frac{\alpha ^4}{p^3 \beta },
\\ & \cZ P (1,0,0,1)=\frac{\alpha ^4}{p^2 \beta ^2},
\  \cZ P (1,0,1,0)=\frac{\alpha ^4 (p+\beta )}{p^3 \beta ^2},
\\  &  \cZ P (1,0,1,1)=\frac{\alpha ^4}{p \beta^3},
\ \cZ P (1,1,0,0)=\frac{\alpha ^4 (p+2 \beta )}{p^3 \beta ^2},
\\ & \cZ P (1,1,0,1)=\frac{\alpha ^4 (p+\beta )}{p^2 \beta ^3},
\\ & \cZ P (1,1,1,0)
=\frac{\alpha ^4 \left(p^2+p\beta +\beta ^2\right)}
           {p^3\beta ^3},
\ \cZ P (1,1,1,1)=\frac{\alpha ^4}{\beta ^4},
\end{align*}
where $\cZ=\cZ(\alpha,\beta,p).$
These were calculated by using the algebraic relation \eqref{alg}.

\end{document}